# Quasars Lensed by Globular Clusters of Spiral and Elliptical Galaxies

Yu. L. Bukhmastova

*Astronomical Institute, St. Petersburg State University, Russia,*
*E-mail: bukh_julia@mail.ru*
*Translated by V. Astakhov*

**Abstract**—Based on the SDSS catalog, we have found new close quasar–galaxy pairs. We analyze the radial distribution of quasars from pairs around galaxies of different types. We show that the quasars from pairs follow the density profile of halo globular clusters. This is new observational evidence that the quasars projected onto the halos of galaxies are magnified by gravitational lensing by halo globular clusters.

INTRODUCTION

This paper is a continuation of the series of works devoted to the problem of quasar–galaxy associations (Baryshev and Ezova 1997; Bukhmastova 2001, 2002, 2003). A pair of objects in which a distant quasar is projected onto the halo of a nearer galaxy will be called a quasar–galaxy association. The halo radius is assumed to be 150 kpc. The catalog of quasars (Veron-Cetty and Veron 1998) and the LEDA database (Paturel 1997) were used previously to select such pairs. A total of 8382 pairs have been selected from 77 483 galaxies and 11 358 quasars. Bukhmastova (2001) pointed out that if the galaxies and quasars were distributed randomly in space, then the number of such pairs would be ~2000. The discrepancy in the number of pairs by more than a factor of 4 provides evidence for the hypothesis about a nonrandom association of quasars with galaxies. Bukhmastova (2001, 2003) and Baryshev and Bukhmastova (2004) considered the following three hypotheses regarding the relatively close association between galaxies and quasars.

1. Among the 8382 pairs found, there are those formed by chance. The increase in the probability of occurrence of random pairs is particularly large where the redshift of a galaxy is too low, since the halo with a radius of 150 kpc in such a galaxy has a large angular area, covering much of the sky. A large number of randomly projected distant objects, including quasars, are seen through such a halo.
2. The pairs are formed through a nonuniform, fractal distribution of galaxies in space. According to his distribution, the space density of galaxies around a selected galaxy decreases as $r^{-a}$, where $r$ is the distance from the galaxy under consideration to the selected galaxy and $a$ is close to unity (Pietronero 1987; Sylos-Labini et al. 1998; Baryshev and Bukhmastiva 2004); i.e., the galaxies tend to be arranged in hierarchical groups. The quasars are currently believed to be active galactic nuclei. Thus, if we observe a galaxy in a pair with a quasar, then there is a high probability that several more active galactic nuclei, i.e., quasars,will be in a pair with this galaxy. Conversely, several galaxies can be in a pair with one quasar. This explains the enhanced probability of occurrence of Arp and Tyson quasar–galaxy associations.
3. Finally, the third main hypothesis of this paper consists in the following. The quasars from associations are distant compact objects (the nuclei of distant galaxies) whose fluxis enhanced by gravitationallenses on the line of sight. Globular clustersin the halos of nearer galaxies and clustered objects of hidden mass can be such lenses. As a result of gravitational lensing, we can see a magnified (and split) image of a distant source. The possible numberof globular clusters in the halos of galaxies and their radial distribution relative to the center of the host galaxy should be taken into account in theoretical estimations of the expected number of such pairs. According to Blakeslee et al. (1997), Blakeslee (1999), and McLaughlin (1999), the globular clusters contain 25% of the baryonic mass of galaxies. In the section entitled "The Expected Number of Pairs", we estimate the expected number of close quasar– galaxy pairs by taking into account the abundance of globular clusters.

The theory of gravitational lensing was well described by Bliokh and Minakov (1989), Zakharov(1997), and Schneider et al. (1992); many papers are devoted to quasar–galaxy associations (see, e.g., Thomas et al. 1995; Zhu Zong-Hong et al. 1997; Benitez et al. 2001; Menard and Bartelmann 2002;Scranton et al. 2005; for the possibility of strong gravitational magnification of quasars, see Yonehara
et al. 2003).

Quasars are observed in pairs with galaxies of various types. If distant quasars and nearer galaxies are associated via halo globular clusters, then the distribution of quasars around elliptical galaxies must follow in a certain way the distribution of globular clusters around these galaxies. Does the distribution of quasars around elliptical and spiral galaxies have peculiarities of its own? This paper is devoted to a discussion of this hypothesis.

## THE QUASAR–GALAXY PAIR SELECTION TECHNIQUE

To select quasar–galaxy pairs, we use the available catalog of 8382 pairs (Bukhmastova 2001) compiled from the LEDA database (Paturel 1997) and the catalog of quasars (Veron-Cetty and Veron 1998). We also select new pairs from the fourth version of the SDSS catalog of quasars and galaxies (www.sdss.org) The selection criteria are the following.
1. We select quasars and galaxies with available data on their spatial coordinates $\alpha$, $\delta$, and $z$. The apparent magnitudes are also known for the quasars.
2. Each quasar in a pair must be farther than the galaxy, i.e., $z_{qso} > z_{gal}$.
3. The quasars must be projected onto the galaxy halos. The halo size does not exceed 150 kpc.
4. We consider nearby quasars with$z$ < 0.3.
5. We select pairs with $z_{gal}/z_{qso}$ > 0.9.

Criterion 5 allows us to reduce significantly the number of random pairs. According to Bukhmastova (2001), the galaxies paired with quasars are either
close to the observer, so that $a = z_{gal}/z_{qso}$ < 0.1, or close to the quasar, so that $a$ > 0.9, avoiding the central location on the observer–quasar ray. There may be a large number of random pairs among those with $a$ < 0.1, because the halo of a nearby galaxy has a large angular size, thereby covering much of the sky. The probability of chance coincidences among the pairs with $a$ > 0.9 is considerably lower. We will call the pairs with $a$ > 0.9 close quasar–galaxy pairs, because the quasar and the galaxy in such a pair are close to one another not only in angular separation, but also in redshift.

Criterion 4 also reduces the sample by removing the distant quasars that can be paired with distant galaxies, which are undetectable in the surveys under consideration.

SELECTION OF PAIRS BASED ON THE LEDA DATABASE AND THE CATALOG OF QUASARS (VERON-CETTY AND VERON 1998)

Bukhmastova (2001) selected 8382 pairs based on criteria 1–3. Given the additional constraints 4 and 5, 64 close quasar–galaxy pairs listed in Table 1 are left. We will call the sample of pairs from Table 1 sample 1. The columns of this table give the following data: 1, the pair number in the original catalog; 2, the quasar name; 3 and 4, the equatorial coordinates of the quasar in radians; 5 and 6, the redshift and apparent magnitude of the quasar; 7, the galaxy name; 8, the galaxy redshift; 9 and 10, the distance from the galaxy center to the quasar projection in kpc and arcsec, respectively; 11, the presumed splitting angle between the quasar images under the assumption that the mass of the gravitational lens is $3*10^5$ $M_\odot$ and 12, the galaxy type according to the NED database (http://nedwww.ipac.caltech.edu).

The splitting angle between the images was calculated using the model of a point lens and corresponds to the radius of the Chwolson–Einstein ring. Although the putative lenses are globular clusters, which are well described by the King model (King 1962), this estimate of the splitting angle is admissible. In particular, for the rays passing outside the core of such a lens, we can use the point lens model, which is simpler for general estimations.

SELECTION OF PAIRS BASED ON THE SDSS CATALOG

Let us select new pairs based on the fourth version the Sloan Digital Sky Survey ([www.sdss.org](www.sdss.org)). 5224 quasars with $z < 0.3$ and 321 516 galaxies are involved in the selection. This original sample of quasars and galaxies was compiled by N.L. Vasil'ev. The result of our selection using criteria 1–5 is presented in Table 2 (sample 2, 64 pairs). The odd rows in the table present data on the quasar in pairs. The columns give the following data: 1, the quasar–galaxy pair number in this sample; 2, the quasar identification number in the SDSS catalog; 3 and 4, the equatorial coordinates of the quasar in radians; 5 and 6, the redshift and apparent magnitude of the quasar; and 9, the presumed splitting angle between the quasar images under the assumption that the mass of the gravitational lens is $3*10^5$ $M_\odot$. The even rows in the table present data on the galaxies in pairs: 2, the galaxy identification number in the SDSS catalog; 3 and 4, the equatorial coordinates of the galaxy in radians; 5 and 6, the redshift and apparent magnitude of the galaxy; 7, the color index of the galaxy; and 8, the distance from the galaxy center to the quasar projection in kpc.

ANALYSIS OF THE SELECTED PAIRS

Let us analyze the derived samples of close quasar–galaxy pairs. Let us determine the linear distances from the galaxy centers at which the quasars from pairs are projected. Figures 1 and 2 show the distributions of quasars from samples 1 and 2, respectively. The distance from the galaxy center to the quasar projection in kpc is along the horizontal axis and the number of quasars is along the vertical axis.

According to the data presented in Table 1, sample 1 of close quasar–galaxy pairs is represented mostly by spiral galaxies. Our main assumption is that the quasars of sample 1 are associated with spiral galaxies via globular clusters in the halos of these galaxies. Thus, Fig. 1 leads us to conclude that 80% of the globular clusters in spiral galaxies are located in their halos at distances up to 40 kpc. Note that the clustered halo objects are predominantly within 10 kpc; further out, the number of such objects decreases sharply. Below, we will establish whether this is actually the case.

To establish the types of galaxies represented in sample 2, let us turn to the paper by Fukugita et al. (1995), who classified the galaxies according to their color indices. Table 3 presents these classifications for the SDSS catalog. Tables 2 and 3 lead us to conclude that about 70% of the galaxies in the pairs of sample 2 are elliptical ones. In the section on globular clusters in the elliptical galaxies A754, A1644, A2124, A2147, A2151, A2152 of the Abell cluster and in the galaxies IC 4051, M49, and M87, we will establish whether this means that the data in Fig. 2 reflect the spatial distribution of globular clusters in elliptical galaxies.

## GLOBULAR CLUSTERS IN SPIRAL GALAXIES AND QUASAR–GALAXY ASSOCIATIONS

To establish the peculiarities of the distribution of globular clusters in the halos of spiral galaxies, we use data on the locations of 150 globular clusters in the MilkyWay halo (Harris 1996). Analysis of these data is presented in Fig. 3. It follows from this analysis that more than 80% of the globular clusters are actually located at distances of no larger than 40 kpc. To determine the density profile of globular clusters in the halos of spiral galaxies in more detail, let us turn to the data on 1164 candidates for globular clusters in the Andromeda Galaxy (Galetti et al. 2004). Figure 4 shows the distribution of these globular clusters in projection onto the galactic plane. The distances from the galaxy center (point 0.0) to the globular clusters are along the axes. We counted the halo globular clusters in rings of a fixed radius. The upper straight line in Fig. 5 represents the density profile of globular clusters in the Andromeda Galaxy. We see that the radial density of globular clusters in the halos of spiral galaxies is well described by a power law in the form

$$n_{GC}(r_p) = A_{0*}r_p^\alpha , \qquad (1)$$

where α = −2.1 ± 0.3 and α = −2.5 ± 0.3 for the Andromeda and the Milky Way, respectively; $n_{GC}$ is the number of globular clusters per unit area of the halo ring; and $r_p = R$ is the galactocentric distance of the globular cluster in projection onto the galactic plane.

Let us select the close pairs from samples 1 and 2 in which the quasar is projected onto the halo of a spiral galaxy. The result is indicated by the lower straight line in Fig. 5. For these quasars, α = −2.6 ±0.3 in a segment up to 80kpc. Figure 5 leads us to conclude that the distribution of globular clusters in the plane of the halos of spiral galaxies and the distribution of quasar projections onto the halos of spiral galaxies are described by a power law with a mean index α ≈−2.4, i.e., the quasars from close quasar–galaxy pairs may follow the halo globular clusters in their radial distribution.

## GLOBULAR CLUSTERS IN ELLIPTICAL GALAXIES AND QUASAR–GALAXY ASSOCIATIONS

Let us select the quasar–galaxy pairs from samples 1 and 2 in which the quasar is projected onto an elliptical galaxy. We will construct a dependence similar to that in Fig. 5. The result indicated by the lower line in Fig. 6 leads us to conclude that the distribution of quasars in the halos of elliptical galaxies obeys a power law with an index α = −1.5±0.3 up to 80 kpc. The quasars father than 80 kpc may be background ones unassociated with the presumed gravitational lenses of the halo.

The upper straight line in Fig. 6 was constructed by analyzing the galaxy-galaxy pairs. We selected the galaxy–galaxy pairs from the SDSS catalog in a similar way as the quasar–galaxy pairs. It was necessary to determine how the radial distribution of galaxies seen through the halo of a nearer galaxy fell off. We see that the galaxies are distributed uniformly in projection onto the sky. Note that the radial distribution of galaxies seen through the halos of nearer galaxies and the distribution of quasars projected onto the halos of elliptical galaxies at distances larger than 80 kpc are identical.

The distribution 263 globular clusters in the elliptical galaxy M49 within 30 kpc of its center is clearly presented in Cote et al. (2003). This galaxy contains a total of ~6000 globular clusters within 100 kpc of its center.

Cote et al. (2001) analyzed the distribution of 278 globular clusters in the elliptical galaxy M87 within 30 kpc of its center. There are ~13 500 globular clusters within 100 kpc of its center. Woodworth and Harris (2000) provided the spatial distribution of globular clusters in the galaxy IC 4051. The question of whether the total area of the halo covered by globular clusters is enough to cover and magnify distant sources (quasars) is considered in the section entitled "The Expected Number of Pairs". The density of globular clusters in elliptical galaxies increases toward the center. As an example, the density profile of globular clusters can be traced for the galaxy A754 (see Fig. 7). The globular clusters in the galaxies A1644, A2124, A2147, A2151, and A2152 show a similar behavior. This led Blakeslee (1999) to conclude that the radial density of globular clusters in the halos of elliptical galaxies is well described by a power law in form (1) with a mean α ≈ −1.5.

Thus, Figs. 6 and 7 suggest that the distribution of globular clusters in the plane of the halos of elliptical galaxies and the distribution of quasar projections onto the halos of elliptical galaxies are described by a power law with an index α ≈ −1.5, i.e., the quasars from close quasar–galaxy pairs follow the halo globular clusters in their radial distribution.

## THE EXPECTED NUMBER OF PAIRS

Let us estimate the expected number of close quasar–galaxy pairs using the above selection criteria, i.e., let us answer the question of how probable is the fact that distant compact sources (galactic nuclei) can be magnified by globular clusters in the halos of nearer galaxies. Without gravitational lensing, such a source would remain invisible for an observer.

The main condition for strong gravitational amplification (by 3–5$^m$) by gravitational lenses with a King mass distribution, which include globular clusters, is the projection of distant source onto the core of such a lens. The Milky Way clusters have core radii in the range 0.08–24 pc (Harris 1996). Let us assume that the parameters of globular clusters in other galaxies are similar to those of the Milky Way clusters and fix the core radius at 9 pc. Assume that the galaxy with the surrounding gravitational lenses lies at $z = 0.08$. The maximum of the distribution of SDSS galaxies lies at this redshift $z$. For definiteness, we take the halo size to be 40 kpc. Most of the halo globular clusters are located within this radius (as follows from the calculations given below, this quantity plays no role). The halo area of such a galaxy is ~ $10^{-4}$ square degrees. If the density of distant sources in the sky is assumed to be $10^6$ objects per square degree, then such a halo will cover ~100 distant sources. Let us now estimate the fraction of the halo area occupied by the cores of globular clusters. To this end, we must estimate the possible number of globular clusters in the halo. Table 4 from Bukhmastova (2003) presents data on the number of globular clusters in spiral and elliptical galaxies. The columns of the table contain the following: 1, the galaxy name; 2, the galaxy type; 3, the number of quasars projected onto the galaxy; and 4, the number of halo globular clusters. Based on these data, we assume that the mean number of globular clusters in the halos of galaxies is $10^3$. Thus, multiplying the number of globular clusters by thearea of the globular cluster core and dividing the value obtained by the area of the galaxy halo, we obtain $5*10^{-5}$, the fraction of the area occupied by globular clusters. Multiplying this value by the number of sources covered by the halo yields $5 *10^{-3}$. This means that five of every thousand galaxies located at $z = 0.08$ may have at least one magnified source (quasar) in their halo. According to Fig. 8, the number of galaxies located at this distance is ~$2.6*10^4$. As a result, the expected number of galaxies in pairs quasars is 130 against 128 galaxies listed in Tables 1 and 2. Thus, these simple estimates are consistent with observations and suggest that close quasar–galaxy pairs can actually be produced by gravitational amplification of distant sources considered as quasars. Gravitational lenses can be globular clusters of galaxy halos or other clustered objects of hidden mass with masses and radii close to those of globular clusters.

## CONCLUSIONS AND OBSERVATIONAL TESTS

Quasars projected onto the halos of nearer galaxies are encountered among the multitude of quasars observed at various distances from us. Among them there are quasars that are close to the galaxies not only in angular separation, but also in redshift. Such quasar–galaxy pairs are called close pairs. In this paper, we developed further the hypothesis that such pairs appear, because the fluxer on the nucleus of the more distant galaxy passes through halo globular clusters of the nearer galaxy, resulting in magnification and splitting of the image of the source that we interpret as a quasar. To corroborate this hypothesis, we analyzed the distribution of quasars in the plane of the halos of these galaxies. The quasars from close pairs were found to follow the density profile of globular clusters in the halos of elliptical and spiral galaxies with slopes of α ≈ −1.5 and α ≈ −2.4 for elliptical and spiral galaxies, respectively. This suggests that quasars do not appear near galaxies by chance and that quasars are associated with galaxies via halo globular clusters.

The quasars from close quasar–galaxy pairs can be observed to study the splitting of their images. The presumed splitting angles between the images are very small (severalmilliarcseconds), but they are nevertheless accessible to such telescopes as the VLTI in Chile. Another observational test consists in the following. If the quasars from pairs are actually the central sources of galaxies magnified by globular clusters in the nearer galaxy, then, since $z_{gal}/z_{qso} > 0.9$ in close pairs, then stars of the host galaxy of the quasar will be mixed with stars of the nearer lensing galaxy for an observer. This may give rise to lines in the galaxy spectra corresponding to two redshifts.

We were unable to perform similar studies with irregular galaxies, because we found no close pairs, i.e., quasars and galaxies close in redshift, among the 203 quasar–irregular galaxy pairs found (Bukhmastova 2001). However, the following fact is of considerable interest: despite the smaller number of quasar–irregular galaxy pairs compared to the number of quasar–elliptical galaxy and quasar–spiral galaxy pairs, the relative contribution of irregular galaxies in pairs is considerably higher, i.e., among the elliptical and spiral galaxies, 1–2% of their total number are in pairs with quasars, while among the irregular galaxies, about 9% of their total number are in pairs with quasars (Bukhmastova 2003). This suggest that there may be an enhanced number of compact objects with masses and radii typical of globular clusters around the irregular galaxies at distances of ∼50−100 kpc. Detection of these effects would be yet another observational tests on the possibility of an association between quasars and nearby galaxies.

## ACKNOWLEDGMENTS

I wish to thank Yu.V. Baryshev for valuable remarks. This work was supported in part by a grant from the President of Russia for support of leading scientific schools (Nsh-8542-2006.2).

## REFERENCES

1. Yu. V. Baryshev and Yu. L. Bukhmastova, *Pis'ma Astron. Zh.* **30**, 493 (2004 [*Astron. Lett.* **30**, 444 (2004)].
2. Yu. V.Baryshev and Yu. L. Ezova, *Astron.Zh.* **74**, 497(1997) [*Astron. Rep.* **41**, 436 (1997)].
3. N. Benitez, J. L. Sanz, and E. Martinez-Gonzalez, *Mon. Not. R. Astron. Soc.* **320**, 241 (2001); astro-ph/0008394.
4. J. P. Blakeslee, *Astron. J.* **118**, 1506 (1999); astro-ph/9906356.
5. J. P. Blakeslee, J. L. Tonry, and M. R. Metzger, *Astron.J.* **114**, 482, (1997).
6. P. V. Bliokh and A. A. Minakov, *Gravitational Lenses* (Naukova Dumka, Kiev, 1989) [in Russian].
7. Yu. L. Bukhmastova, *Astron. Zh.* **78**, 675 (2001) [*Astron.Rep.* **45**, 581 (2001)].
8. Yu. L. Bukhmastova, *Astrofizika* **45**, 231 (2002) [*Astrophys.* **45**, 191 (2002)].
9. Yu. L. Bukhmastova, *Pis'ma Astron. Zh.* **29**, 253(2003) [*Astron. Lett.* **29**, 214 (2003)].
10. P. Cote, D. E. McLaughlin, D. A. Hanes, et al., *Astrophys.J.* **559**, 828 (2001); astro-ph/0106005.
11. P. Cote, D. E. McLaughlin, J. G. Cohen, and J. P.Blakeslee, *Astrophys. J.* **591**, 850 (2003); astro-ph/0303229.
12. M. Fukugita, K. Shimasaka, and T. Ichikawa, *Publ.Astron. Soc. Pac.* **107**, 945 (1995).
13. S. Galleti, L. Federici, M. Bellazzini, et al., *Astron.Astrophys.* **416**, 917 (2004); ftp://cdsarc.ustrasbg.fr/pub/cats/J/A+A/416/917;http://cdsweb.u-strasbg.fr/viz-bin/VizieR?-source=J/A+A/416/917.
14. W. E. Harris, *Astron. J.* **112**, 1487 (1996); www.physics.mcmaster.ca/Globular.html.

15. I. King, *Astron. J.* **67**, 471 (1962).
16. D. E. McLaughlin, *Astron. J.* **117**, 2398 (1999).
17. B. Menard and M. Bartelmann, *Astron. Astrophys.* **386**, 784 (2002); astro-ph/0203163.
18. J. Paturel, *Astrophys. J., Suppl. Ser.* **124**, 109 (1987).
19. L. Pietronero, *Physica A* **144**, 257 (1997).
20. P. Schneider, J. Ehlers, and E. E. Falko, *Gravitational Lenses* (Springer-Verlag, New York, 1992).
21. R. Scranton, B. Menard, G. Richards, et al., *Astroph. J.* **633**, 589 (2005); astro-ph/0504510.
22. F. Sylos-Labini, M. Montuori, and L. Pietronero, *Phys. Rep.* **293**, 61 (1998).
23. P. A. Thomas, R. L. Webster, and M. J. Drinkwater, *Astron. Astrophys.* **299**, 353 (1995); astro-ph/9408088.
24. M. P. Veron-Cetty and P. Veron, *Quasars and ActiveGalactic Nuclei ESO Sci. Rep. 18* (1998).
25. S. C. Woodworth and W. E. Harris, *Astron. J.* **119**, 2699 (2000); astro-ph/0002292.
26. A. Yonehara, M. Umemura, and H. Susa, *Astron. Soc. Pac.* **55**, 1059 (2003); astro-ph/0310296.
27. A. F. Zakharov, *Gravitational Lenses and Microlenses* (Yanus-K, Moscow, 1997) [in Russian].
28. Zhu Zong-Hong, Wu Xiang-Ping, and Fang Li-Zhi, *Astrophys. J.* **490**, 31 (1997); astro-ph/9706289.

# TABLES

**Table 1.** Close quasar–galaxy pairs found based on the catalog of quasars (Veron-Cetty and Veron 1998) and the LEDA database (Paturel 1997)

| N | Name_qso | RA_qso | DEC_qso | z_qso | m_qso | Name_gal | z_gal | r_кпк | r" | Q" | type_gal |
|---|---|---|---|---|---|---|---|---|---|---|---|
| 107 | PG 0026+129 | 0.116217 | 0.22674 | 0.142 | 15.41 | PGC001790 | 0.1419 | 0.7 | 0.2 | 0.0020 | Sy1 |
| 743 | Q 0051-3933 | 0.226049 | -0.69031 | 0.224 | 17.3 | LEDA0125101 | 0.2233 | 12.5 | 2.3 | 0.0021 | - |
| 783 | PG 0052+251 | 0.227700 | 0.43906 | 0.155 | 15.43 | PGC003237 | 0.1549 | 0.4 | 0.1 | 0.0017 | Sb,Sy1 |
| 1764 | F 9 | 0.357152 | -1.03090 | 0.046 | 13.83 | PGC005109 | 0.045 | 34.1 | 31.2 | 0.0611 | Sb,Sy1 |
| 1885 | MS 01360-5614 | 0.418995 | -0.98160 | 0.086 | 15.5 | LEDA0138429 | 0.0859 | 24.1 | 11.5 | 0.0054 | - |
| 1952 | MARK 1014 | 0.511694 | 0.00266 | 0.163 | 15.69 | PGC007551 | 0.1628 | 0.5 | 0.1 | 0.0003 | Sb,Sy1 |
| 1961 | RXS J02070+2930 | 0.541692 | 0.51094 | 0.11 | 16.0 | PGC008076 | 0.1094 | 51.8 | 19.5 | 0.0081 | Sbc,Sy1 |
| 2019 | 4U 0241+61 | 0.702538 | 1.08659 | 0.045 | 12.19 | LEDA0074110 | 0.044 | 7.5 | 7.0 | 0.0639 | Sy1 |
| 2044 | MS 02448+1928 | 0.719315 | 0.33989 | 0.176 | 16.66 | LEDA0138537 | 0.1759 | 29.0 | 6.8 | 0.0013 | Sy1 |
| 2077 | PKS 0312-77 | 0.841808 | -1.34478 | 0.225 | 16.1 | LEDA0088074 | 0.2228 | 8.0 | 1.4 | 0.0037 | Sy1 |
| 2078 | ESO 199-IG23 | 0.833351 | -0.89939 | 0.078 | 15.1 | PGC011951 | 0.0779 | 16.4 | 8.6 | 0.0066 | Sc,S0/a |
| 2091 | 0321-424 | 0.880528 | -0.74014 | 0.2 | 16.6 | LEDA0088125 | 0.1999 | 34.2 | 7.1 | 0.0010 | Sy1 |
| 2253 | MS 03574+1046 | 1.036063 | 0.18813 | 0.182 | 16.78 | LEDA0138680 | 0.1814 | 16.0 | 3.6 | 0.0030 | Sy1 |
| 2278 | IRAS 04505-2958 | 1.267770 | -0.52316 | 0.286 | 16.0 | LEDA0075249 | 0.2858 | 35.2 | 5.0 | 0.0007 | Sy1 |
| 2291 | IRAS 06115-3240 | 1.620967 | -0.57043 | 0.05 | 14.1 | PGC018655 | 0.0499 | 72.7 | 60.1 | 0.0008 | Sab,Sy2 |
| 2302 | VII Zw 118 | 1.843074 | 1.12883 | 0.079 | 14.61 | PGC020174 | 0.0788 | 0.8 | 0.4 | 0.0091 | Sab |
| 2330 | IRAS 07483+0328 | 2.043739 | 0.06061 | 0.099 | 15.2 | LEDA0097223 | 0.0989 | 7.1 | 2.9 | 0.0041 | Sy1 |
| 2332 | B3 0754+394 | 2.071007 | 0.68899 | 0.096 | 14.36 | LEDA0139042 | 0.0957 | 7.3 | 3.1 | 0.0075 | Sy1.5 |
| 2344 | MS 08019+2129 | 2.102874 | 0.37507 | 0.118 | 15.92 | LEDA0139051 | 0.1179 | 29.8 | 10.4 | 0.0029 | - |
| 2349 | PG 0804+761 | 2.114422 | 1.32980 | 0.1 | 14.71 | PGC022946 | 0.0999 | 2.6 | 1.1 | 0.0040 | Sy1 |
| 2374 | PG 0844+349 | 2.288846 | 0.60974 | 0.064 | 14.5 | PGC024702 | 0.0639 | 0.2 | 0.1 | 0.0006 | Sy1 |
| 2396 | MS 09063+1111 | 2.383661 | 0.19540 | 0.16 | 16.93 | LEDA0139171 | 0.1599 | 138 | 35.6 | 0.0016 | Sy |
| 2467 | IRAS 09435-1307 | 2.546415 | -0.22897 | 0.131 | 16.3 | LEDA0082528 | 0.1309 | 6.3 | 1.9 | 0.0023 | Sy2 |
| 2469 | MS 09441+1333 | 2.548915 | 0.23671 | 0.131 | 16.05 | LEDA0139211 | 0.1309 | 48.3 | 15.2 | 0.0023 | Sy1 |
| 2547 | PG 1001+05 | 2.625506 | 0.09529 | 0.161 | 16.23 | PGC029208 | 0.1609 | 1.0 | 0.2 | 0.0015 | Sy1 |
| 2617 | MS 10302-2757 | 2.750042 | -0.48810 | 0.148 | 16.0 | LEDA0139395 | 0.1479 | 101.9 | 28.4 | 0.0018 | - |
| 2708 | CSO 292 | 2.827862 | 0.61073 | 0.147 | 16.6 | LEDA0139451 | 0.1469 | 25.1 | 7.0 | 0.0019 | Sy1 |
| 2788 | PG 1114+445 | 2.942334 | 0.77665 | 0.144 | 16.05 | PGC034449 | 0.1439 | 23.7 | 6.7 | 0.0019 | Sy1 |
| 2809 | PG 1115+407 | 2.948588 | 0.71044 | 0.154 | 16.02 | PGC034570 | 0.1539 | 20.2 | 5.4 | 0.0017 | Sy1 |
| 2863 | WAS 26 | 3.048508 | 0.38775 | 0.063 | 14.9 | PGC036264 | 0.0628 | 91.1 | 59.8 | 0.0143 | Sy1 |

| | | | | | | | | | | |
|---|---|---|---|---|---|---|---|---|---|---|
| 3490 | PG 1211+143 | 3.192854 | 0.25012 | 0.085 | 14.19 | PGC039086 | 0.0849 | 2.8 | 1.3 | 0.0055 | Sb |
| 7531 | PG 1307+085 | 3.435113 | 0.15003 | 0.155 | 15.11 | PGC045656 | 0.1549 | 0.6 | 0.1 | 0.0017 | Sy1 |
| 7829 | PG 1351+64 | 3.629281 | 1.11715 | 0.088 | 14.28 | PGC049340 | 0.0878 | 2.4 | 1.1 | 0.0073 | Sy1 |
| 7867 | MS 13591+0430 | 3.661228 | 0.07879 | 0.163 | 16.84 | LEDA0140243 | 0.1629 | 60.4 | 15.2 | 0.0015 | - |
| 7894 | PG 1402+261 | 3.678230 | 0.45665 | 0.164 | 15.54 | PGC050237 | 0.1639 | 38.2 | 9.6 | 0.0015 | SBb,Sy1 |
| 7895 | PG 1404+226 | 3.682826 | 0.39504 | 0.098 | 15.82 | PGC050313 | 0.0979 | 13.0 | 5.4 | 0.0042 | Sc |
| 7945 | PG 1411+442 | 3.716824 | 0.77207 | 0.089 | 14.99 | PGC050824 | 0.0889 | 2.3 | 1.1 | 0.0051 | Sy1 |
| 7968 | PG 1415+451 | 3.730968 | 0.78828 | 0.114 | 15.74 | PGC051016 | 0.1139 | 6.8 | 2.4 | 0.0031 | Sy1 |
| 7983 | PG 1416-129 | 3.736553 | -0.22601 | 0.129 | 16.1 | PGC051142 | 0.1289 | 1.1 | 0.3 | 0.0024 | Sy1 |
| 8030 | PG 1435-067 | 3.820620 | -0.11792 | 0.129 | 16.01 | PGC052314 | 0.1289 | 11.3 | 3.6 | 0.0024 | Sy1 |
| 8058 | 3C 305.0 | 3.875889 | 1.10787 | 0.042 | 13.74 | PGC052924 | 0.0414 | 3.1 | 3.0 | 0.0563 | SB0 |
| 8060 | PG 1448+273 | 3.878870 | 0.47756 | 0.065 | 15.01 | PGC053030 | 0.0648 | 7.9 | 5.0 | 0.0134 | Sa,Sbc |
| 8081 | PG 1519+226 | 4.010024 | 0.39516 | 0.137 | 16.09 | PGC054782 | 0.1369 | 24.7 | 7.4 | 0.0021 | Sy1 |
| 8098 | PG 1552+085 | 4.155279 | 0.14867 | 0.119 | 16.02 | PGC056356 | 0.1189 | 0.4 | 0.1 | 0.0028 | - |
| 8116 | MARK 876 | 4.248145 | 1.14918 | 0.129 | 15.49 | PGC057553 | 0.128 | 0.2 | 0.1 | 0.0077 | Sb |
| 8117 | TON 256 | 4.241782 | 0.45720 | 0.131 | 15.41 | PGC057571 | 0.1309 | 0.5 | 0.1 | 0.0023 | Sy1.5 |
| 8122 | 3C 332.0 | 4.257665 | 0.56718 | 0.152 | 16.0 | PGC057750 | 0.1513 | 7.9 | 2.1 | 0.0046 | Sy1 |
| 8123 | MARK 877 | 4.267075 | 0.30589 | 0.114 | 15.39 | PGC057859 | 0.1114 | 4.5 | 1.6 | 0.0161 | Sy1 |
| 8142 | MS 17108+1624 | 4.497909 | 0.28648 | 0.187 | 17.19 | LEDA0140750 | 0.1869 | 91.6 | 20.1 | 0.0011 | - |
| 8158 | IRAS 17500+5046 | 4.669090 | 0.88613 | 0.3 | 15.4 | LEDA0097548 | 0.2995 | 26.0 | 3.5 | 0.0010 | Sy1 |
| 8159 | IRAS 17596+4221 | 4.710658 | 0.73935 | 0.053 | 14.5 | PGC061280 | 0.0527 | 1.2 | 0.9 | 0.0248 | Sa |
| 8166 | MS 20078-3622 | 5.270428 | -0.63475 | 0.177 | 16.5 | LEDA0140881 | 0.1769 | 9.3 | 2.1 | 0.0013 | - |
| 8168 | MARK 509 | 5.416789 | -0.19032 | 0.035 | 13.12 | PGC065282 | 0.0342 | 2.8 | 3.4 | 0.0945 | Sb |
| 8172 | ESO 235-IG26 | 5.478733 | -0.91106 | 0.051 | 13.6 | PGC065837 | 0.0509 | 1.5 | 1.2 | 0.0154 | Sc |
| 8183 | IRAS 21219-1757 | 5.593365 | -0.31349 | 0.113 | 14.5 | PGC066703 | 0.1121 | 7.5 | 2.7 | 0.0095 | Sy1 |
| 8191 | II Zw 136 | 5.628774 | 0.17308 | 0.063 | 14.64 | PGC066930 | 0.0629 | 6.3 | 4.1 | 0.0101 | Sa |
| 8224 | PB 5041 | 5.754975 | 0.02194 | 0.16 | 16.4 | LEDA0140992 | 0.1599 | 13.8 | 3.5 | 0.0016 | Sy1 |
| 8231 | MS 21595-5713 | 5.757433 | -0.99895 | 0.083 | 15.5 | LEDA0089182 | 0.0829 | 26.2 | 13.0 | 0.0058 | - |
| 8248 | MARK 304 | 5.824003 | 0.24415 | 0.067 | 14.66 | PGC068493 | 0.0657 | 17.6 | 11.0 | 0.0328 | Sab,E |
| 8279 | MR 2251-178 | 5.983999 | -0.31151 | 0.068 | 14.36 | PGC069953 | 0.0651 | 57.1 | 36.1 | 0.0487 | Sy1 |
| 8313 | 3C 459.0 | 6.082639 | 0.06658 | 0.22 | 16.68 | PGC070899 | 0.2197 | 18.3 | 3.4 | 0.0014 | - |
| 8332 | MS 23182-4220 | 6.101147 | -0.73884 | 0.212 | 17.0 | LEDA0089418 | 0.2118 | 102.3 | 19.9 | 0.0013 | - |
| 8356 | MS 23409-1511 | 6.199802 | -0.26533 | 0.137 | 15.91 | LEDA0141129 | 0.1369 | 74.4 | 22.4 | 0.0021 | NLSy1 |
| 8366 | PKS 2349-01 | 6.236810 | -0.02498 | 0.174 | 15.33 | PGC072664 | 0.1665 | 17.7 | 4.38 | 0.0120 | Sb |

**Table 2.** Close quasar–galaxy pairs found based on the fourth version of the SDSS catalog (www.sdss.org)

```
N   name_QSO              RA_QSO     DEC_QSO    Z_QSO    m_r_QSO                    Q"
    name_gal              RA_gal     DEC_gal    Z_gal    m_r_gal    g'-r'    r
--------------------------------------------------------------------------------
1   75094095038513152     2.549328   0.003205   0.12670  17.928                    0.000624
    135331280890888190    2.549241   0.003356   0.12631  16.695     0.888   86.7
2   756570579187779392    2.616505   0.010240   0.06557  17.954                    0.001208
    756570579481395520    2.616020   0.010227   0.06518  17.133     0.832  139.3
3   776285722226196948    2.813987   0.016559   0.11498  17.770                    0.000330
    142649771201396740    2.813991   0.016638   0.11489  17.354     0.926   36.5
4   82132249197150208     3.330977   0.017590   0.08956  17.909                    0.003614
    82132249033572352     3.330766   0.017277   0.08346  16.376     0.704  134.2
5   110279118540505090    0.128017  -0.003049   0.06044  18.187                    0.002622
    110279118364344320    0.128422  -0.003222   0.05891  16.499     0.801  115.7
6   110279118540505090    0.128017  -0.003049   0.06044  18.187                    0.002000
    110279118506950660    0.127606  -0.003289   0.05954  15.304     0.846  126.2
7   113093893557321730    0.465772  -0.011845   0.08266  17.881                    0.000933
    113375891036473330    0.465796  -0.011786   0.08229  16.585     0.860   22.4
8   116192138483466240    0.814456   0.001082   0.10734  17.241                    0.001293
    115908768071417860    0.814697   0.001210   0.10615  17.463     0.881  118.4
9   126323700604600320    2.335550   0.995266   0.14364  18.647                    0.000927
    136175426522316800    2.335669   0.995051   0.14254  16.512     1.058  122.6
10  126886685093920770    2.399436   0.989773   0.11136  17.409                    0.001141
    126886685072949250    2.399617   0.989458   0.11036  16.796     0.947  147.8
11  139271612095528960    3.283945   1.159152   0.04679  16.292                    0.002662
    139271612091334660    3.282914   1.159368   0.04584  16.858     0.626   97.6
12  143495708966649860    2.902918   0.015554   0.09752  17.865                    0.001529
    78191514232029184     2.903113   0.015500   0.09615  15.753     0.883   80.9
13  151938426139049980    3.949294   0.044272   0.12155  18.329                    0.001628
    166012336379789310    3.949295   0.044041   0.11915  17.612     0.874  109.9
14  161791502073200640    2.693336   0.090672   0.18952  18.102                    0.000683
    280857043101810690    2.693403   0.090765   0.18848  17.158     1.160   76.4
15  169391156742848510    3.382785   1.114007   0.12839  18.655                    0.000856
    220338146708029440    3.383221   1.114201   0.12764  17.084     1.057  137.1
16  448617234042978300    3.392478   1.108620   0.13353  17.980                    0.001258
```

```
      169391156516356100 3.392358  1.108423 0.13179 17.204 0.976 105.1
17  448617234042978300 3.392478  1.108620 0.13353 17.980                 0.000936
      169391156533133310 3.392817  1.108815 0.13256 17.305 0.970 127.5
18 2020429073251041305.767846 -0.149729 0.21141 17.749                 0.001208
      201760293615828990 5.768045 -0.149702 0.20741 17.445 1.351 141.3
19  203730612249427970 5.981513 -0.155610 0.06489 18.204                 0.001283
      204012313970212860 5.981939 -0.155371 0.06446 17.589 0.686 137.7
20  204293807166980100 6.059609 -0.160396 0.19287 17.096                 0.001073
      204575143651966980 6.059414 -0.160342 0.19023 17.349 0.913 134.2
21  214708269556432900 2.268938  0.778892 0.12539 17.738                 0.001983
      214989756038119420 2.268786  0.778702 0.12164 17.422 1.061 105.7
22  215552788279590910 2.389240  0.881362 0.14935 18.354                 0.000657
      215834232822431740 2.388972  0.881551 0.14875 16.451 1.096 143.4
23  229627051110301700 4.303068  0.747370 0.22831 18.614                 0.000810
      229908548753031170 4.303174  0.747535 0.22619 17.071 1.327 137.2
24  229908547347939330 4.331700  0.713217 0.15527 17.781                 0.000182
      230190032479059970 4.331563  0.713410 0.15522 17.741 1.086 127.4
25  233004486228967420 2.229199  0.651022 0.09881 18.092                 0.000256
      233004486224773120 2.228900  0.650873 0.09877 17.378 0.903 114.8
26  246515465193521150 2.789845  0.889340 0.15390 19.105                 0.000218
      284234355952320510 2.789781  0.889311 0.15383 17.697 1.124  28.7
27  265094338734718980 2.576772  0.773432 0.01532 15.941                 0.005936
      265375787199234050 2.576874  0.773590 0.01482 15.119 0.581  12.5
28  276915144645148670 5.461538  0.015008 0.10619 17.160                 0.000430
      276633769497067520 5.461480  0.015130 0.10606 16.583 1.006  58.6
29  282545493618196480 2.903022  0.120184 0.23546 19.521                 0.000549
      282827321495257090 2.902927  0.120228 0.23442 16.553 1.437  80.1
30  293523379602849790 3.559888  0.955520 0.10736 17.079                 0.000900
      293523379409911810 3.559393  0.955705 0.10678 17.212 0.834 148.3
31  329833770456711170 4.292484  0.717022 0.03352 16.441                 0.002950
      329833770637066240 4.292085  0.717076 0.03292 15.721 0.841  75.6
32  343062906536984580 2.651629  0.748200 0.07182 17.437                 0.001145
      343062906713145340 2.651827  0.747892 0.07140 17.467 0.938 105.9
33  380499427330097150 3.934319  0.661598 0.14887 19.104                 0.001219
      380781025917140990 3.934093  0.661693 0.14684 16.845 1.046 112.8
34  386411569250041860 3.280641  0.786771 0.06220 17.233                 0.000288
      386129060209623040 3.280745  0.786733 0.06218 17.266 0.909  22.8
35  390633933718421500 4.021163  0.560438 0.11171 18.644                 0.000589
      390633933710032900 4.021067  0.560164 0.11144 17.201 1.023 128.9
36  390633933718421500 4.021163  0.560438 0.11171 18.644                 0.001716
      390633933714227200 4.020835  0.560401 0.10946 17.527 0.985 124.6
37  401892407924228100 2.725796  0.655738 0.09867 18.173                 0.000715
      402173892438786050 2.725749  0.655564 0.09836 15.675 0.913  72.5
38  403299815108116480 2.795763  0.697314 0.13812 18.035                 0.001452
      403581426554896380 2.795686  0.697285 0.13565 17.105 1.028  34.6
39  448617234042978300 2.480281  0.611542 0.13353 18.274                 0.000760
      358826732175753220 2.480169  0.611648 0.13289 17.352 1.088  72.6
40  471980327439433730 3.915305  0.797431 0.12184 17.862                 0.000476
      471980327422656510 3.915627  0.797254 0.12163 17.201 0.791 138.4
41  473387807998476290 4.190302  0.650460 0.06809 17.805                 0.001358
      473387807939756030 4.190645  0.650464 0.06756 16.011 0.851  80.9
42   92548344252989440 3.070469 -0.030420 0.10522 16.922                 0.000886
       92828847581429760 3.070388 -0.030341 0.10468 17.112 0.678  48.5
43  126886685253304320 2.407567  0.989370 0.11925 17.704                 0.003586
      127168159089164290 2.407900  0.989299 0.10882 16.997 0.704  86.8
44  142088085100822530 2.719842  0.028602 0.09848 18.275                 0.000693
      142088085113405440 2.719963  0.028921 0.09819 17.526 0.736 138.9
45  143495708966649860 2.902918  0.015554 0.09752 17.865                 0.002282
      143495708941484030 2.903174  0.015400 0.09452 16.269 0.589 117.8
46  187967499967386620 0.568068 -0.158125 0.04136 16.537                 0.001919
      187967499925443580 0.567524 -0.158094 0.04097 16.280 0.669 101.7
47  203730612249427970 5.981513 -0.155610 0.06489 18.204                 0.000338
      203730612245233660 5.981918 -0.155387 0.06486 16.517 0.576 131.0
48  265094338734718980 2.576772  0.773432 0.01532 15.941                 0.002026
      265094338780856320 2.575170  0.773095 0.01526 13.766 0.492  88.5
49  279449281956413440 2.493193  0.091267 0.08539 17.387                 0.000915
      279449281968996350 2.493545  0.091066 0.08501 17.232 0.727 145.9
50  281138587087405060 2.722903  0.113368 0.04401 17.392                 0.001152
      281138587099987970 2.722698  0.113192 0.04385 14.828 0.633  54.2
51  286205176245649410 3.079429  0.973892 0.05349 17.928                 0.001697
      286205176061100030 3.078680  0.974266 0.05298 17.295 0.556 134.5
52  327863107453976580 3.895659  0.932504 0.09706 17.389                 0.001207
      327581609760915460 3.895738  0.932300 0.09621 17.657 0.640  83.8
53  327863107453976580 3.895659  0.932504 0.09706 17.389                 0.001625
      327581609786081280 3.895284  0.932804 0.09553 16.624 0.665 148.9
```

```
54 341374031963881470 2.407220  0.646374  0.08641 17.642                0.000567
   341656661066252290 2.407325  0.646287  0.08626 15.724  0.502  44.2
55 357700815143567360 2.333050  0.540445  0.19697 17.811                0.000752
   357982217877585920 2.333077  0.540458  0.19561 17.673  0.988  18.2
56 369240320803602430 3.091583  1.013120  0.19132 18.562                0.000454
   369803377162321920 3.091289  1.013123  0.19085 17.446  0.822 104.7
57 394574696747630590 4.098871  0.631138  0.06755 17.795                0.002750
   394855002608238590 4.098653  0.630982  0.06546 16.170  0.484  67.8
58 402736861061054460 2.787675  0.687035  0.05472 18.214                0.000866
   403299813640110080 2.787869  0.687066  0.05458 16.504  0.663  37.6
59 410055694239137790 3.343186  0.717970  0.06691 16.947                0.002931
   410055694213971970 3.342752  0.717642  0.06459 16.798  0.545 131.9
60 112249472834076670 0.357779 -0.018153  0.05431 16.289                0.001017
   112249472808910850 0.357892 -0.017646  0.05412 17.460  0.333 126.5
61 113938422875291650 0.530331  0.006949  0.07800 18.337                0.001752
   113938422900457470 0.530044  0.006862  0.07685 16.367  0.376  99.4
62 183182688511852540 6.269681 -0.179214  0.07376 16.426                0.001229
   182901461552726020 6.269695 -0.179224  0.07325 17.115  0.159   5.4
63 236100994553872380 3.064133  0.098798  0.10204 17.504                0.001099
   236382447577595900 3.064365  0.098753  0.10126 16.828  0.367  98.2
64 383315279360294910 2.788191  0.727263  0.00239 17.531                0.007516
   383033745986289660 2.788458  0.727128  0.00237 17.297 -0.111   2.8
```

**Table 3.** Classification of SDSS galaxies according to their color indices (Fukugita et al. 1995)

| Type galaxy | $g'-r'$ | $g'-r'$ z=0.2 |
|---|---|---|
| E | 0.77 | 1.31 |
| S0 | 0.68 | 1.13 |
| Sab | 0.66 | 1.02 |
| Sbc | 0.52 | 0.71 |
| Scd | 0.48 | 0.62 |
| Im | 0.20 | 0.32 |

**Table 4.** Estimated number of globular clusters in spiral and elliptical galaxies

| PGC041297 | E2 | 3 | есть |
|---|---|---|---|
| PGC044324 | Sab | 1 | $10^3$-$10^4$ |
| PGC043008 | E1-2 | 1 | $10^3$-$10^4$ |
| PGC044553 | E | 1 | $10^3$-$10^4$ |
| PGC013344 | S0 | 2 | 385±80 |
| PGC013418 | E1pec | 1 | 5340±1780 |
| PGC013433 | E1 | 1 | 950±140 |
| PGC024930 | Sb | 4 | 310±100 |
| PGC032226 | E5 | 1 | 210±50 |
| PGC032256 | E1 | 6 | 260±140 |
| PGC036487 | E3 | 2 | 14000±2500 |
| PGC039764 | E1 | 2 | 1050±120 |
| PGC039246 | Sb | 129 | 620±310 |
| PGC041327 | E0 | 2 | 13000±500 |
| PGC041968 | E0 | 34 | 2400 |
| PGC042051 | E6 | 3 | 1000±300 |
| PGC042628 | E5 | 7 | 1900±400 |
| PGC000218 | Sab | 1 | 500±160 |

FIGURES

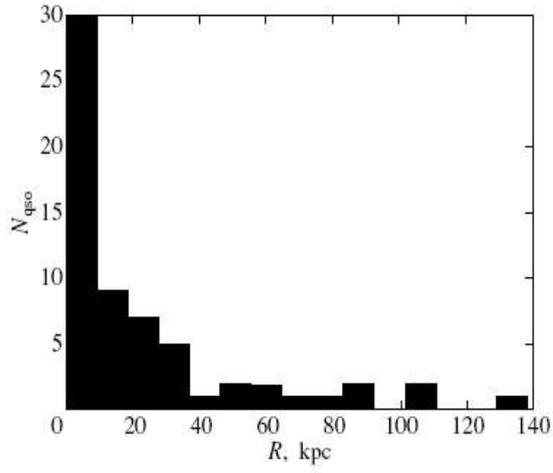

Fig. 1. Distribution of 64 quasars from sample 1 relative to the centers of the galaxies onto the halos of which they are projected.

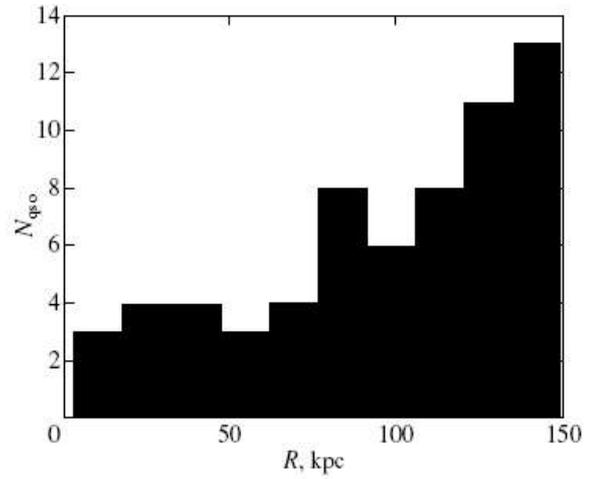

Fig. 2. Same as Fig. 1 for sample 2.

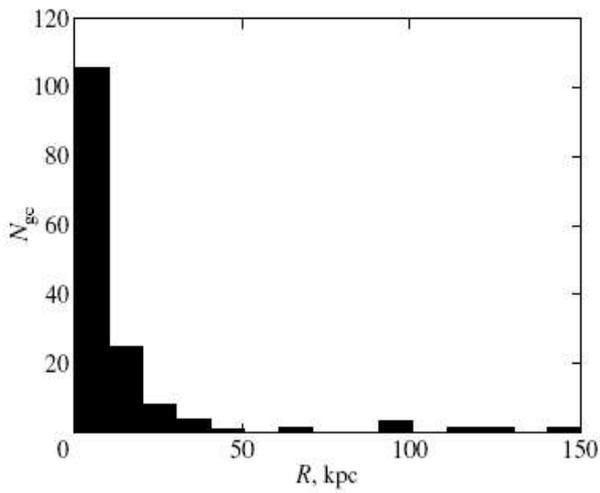

Fig. 3. Distribution of 150 Milky Way globular clusters relative to the Galactic center.

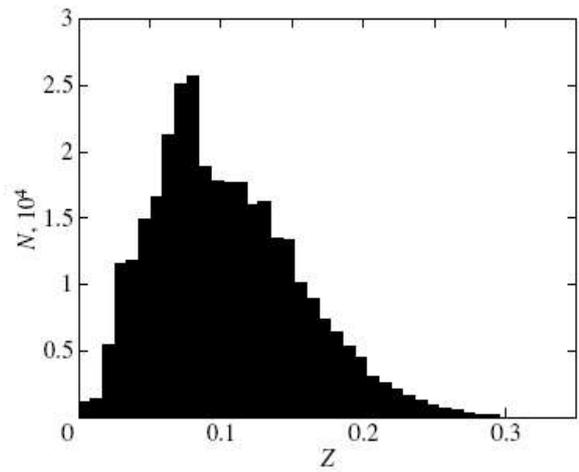

Fig. 8. Redshift distribution of SDSS galaxies.

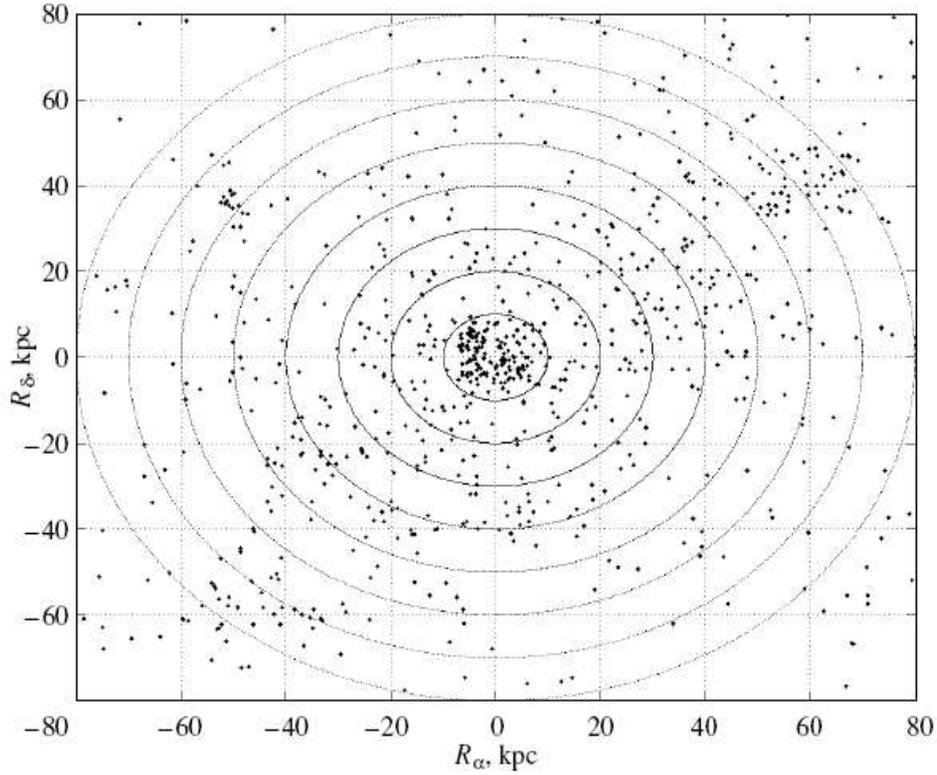

**Fig. 4.** Distribution of globular clusters of the Andromeda Galaxy in the halo plane. The equatorial coordinates recalculated to galactocentric distances are along the axes.

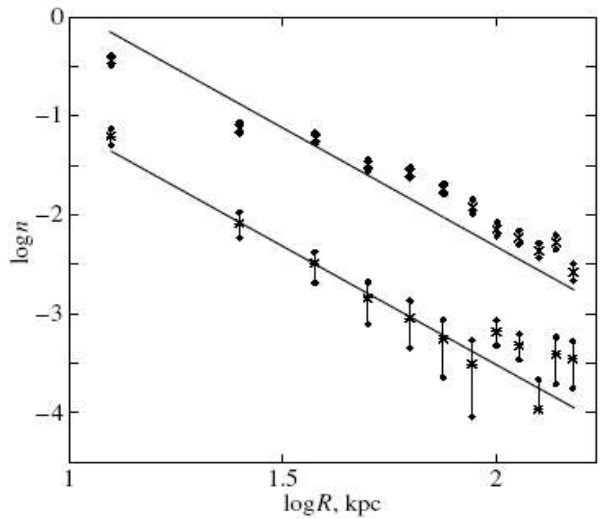

**Fig. 5.** Density profiles of globular clusters in the Andromeda Galaxy (upper straight line) and quasars (lower straight line) projected onto the halos of spiral galaxies. The logarithm of the galactocentric distance of each object (in kpc) is along the horizontal axis; the logarithm of the number of objects (globular clusters and quasars) per unit area of the halo ring is along the vertical axis. The distribution of globular clusters in the plane of the halos of spiral galaxies and the distribution of quasar projections onto the halos of spiral galaxies are well described by a power law with an index $\alpha \approx -2.4$.

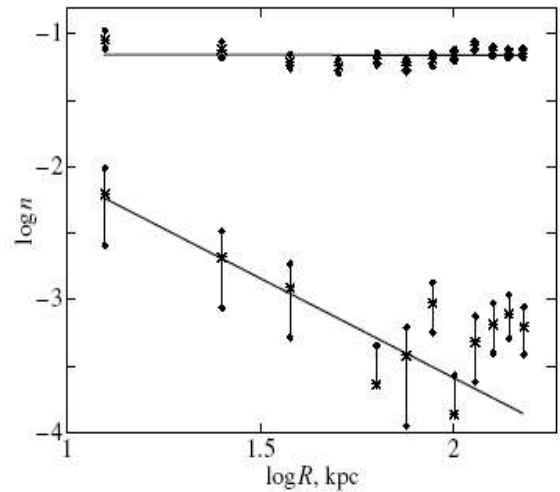

**Fig. 6.** Radial distribution of galaxies seen through the halos of nearer galaxies (upper straight line). The density profile of quasars projected onto the halos of elliptical galaxies (lower straight line). The logarithm of the galactocentric distance of each object (2 corresponds to 100 kpc) is along the horizontal axis; the logarithm of the number of objects per unit area of the halo ring is along the vertical axis. The distribution of quasar projections onto the halos of elliptical galaxies is well described by a power law with an index $\alpha \approx -1.5$.

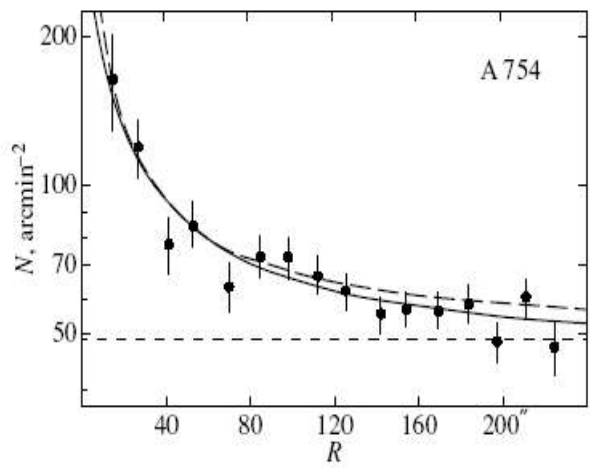

**Fig. 7.** Density profile of globular clusters in the elliptical galaxy A754 (Blakeslee 1999). The galactocentric distances of the globular clusters (in arcsec) are along the horizontal axis; the number of globular clusters per unit area of the halo is along the vertical axis.